\documentclass[sigconf]{acmart}

\renewcommand\footnotetextcopyrightpermission[1]{} 
\def\BibTeX{{\rm B\kern-.05em{\sc i\kern-.025em b}\kern-.08emT\kern-.1667em\lower.7ex\hbox{E}\kern-.125emX}}
    
\usepackage{subcaption}
\usepackage[algoruled,boxed,lined]{algorithm2e}

\usepackage[disable,textsize=scriptsize,backgroundcolor=green]{todonotes}
\newcommand\ad[1]{\todo{{\bf AD}: #1}}

\begin{document}

%
\title{A Conceptual Framework for Evaluating Fairness in Search}

%
\author{Anubrata Das}
\email{anubrata@utexas.edu}
\affiliation{%
  \institution{University of Texas at Austin}
  \city{Austin}
  \state{Texas}
  \country{USA}
}
\author{Matthew Lease}
\email{ml@utexas.edu}
\affiliation{%
  \institution{University of Texas at Austin}
  \city{Austin}
  \state{Texas}
  \country{USA}
}

%
\renewcommand{\shortauthors}{Das and Lease}

%
\begin{abstract}
While search efficacy has been evaluated traditionally on the basis of result relevance, fairness of search has attracted recent attention. In this work, we define a notion of {\em distributional fairness} and provide a conceptual framework for evaluating search results based on it. As part of this, we formulate a set of axioms which an ideal evaluation framework should satisfy for distributional fairness. We show how existing TREC test collections can be repurposed to study fairness, and we measure potential data bias to inform test collection design for fair search. A set of analyses show metric divergence between relevance and fairness, and we describe a simple but flexible interpolation strategy for integrating relevance and fairness into a single metric for optimization and evaluation. 

\end{abstract}

%
%




%

%

%
\maketitle

\section{Introduction}

Algorithmic fairness is now receiving significant attention, and with search systems increasingly mediating human information access, it is recognized that search systems must be fair as well as accurate. However, while the idea of fairness is intuitive, there are many competing definitions of how to operationalize it in practice.

There is also increasing recognition today that dataset bias (e.g., imbalance) can lead to biased training or evaluation. For example, while one might desire balanced search results, an imbalanced dataset distribution can make this goal more difficult to achieve in practice. To the extent relevant information is more scarce for some perspectives or categories, imbalance in relevant information may lead to imbalance in retrieved results. When relevant information for given a category is scarce (or completely absent), it may be difficult (or impossible) for IR systems to find any relevant results from that category to retrieve. Diversity thus plays a role in test collection design, and decisions must be made as to which categories it will be important to ensure diversity in terms of documents included in the collection and relevance annotation.

In this work, we define a notion of {\em distributional fairness} and provide a conceptual framework for evaluating search results based on it (Section~\ref{sec:approach}). Section~\ref{subsec:desiderata} formulates a set of axioms which an ideal evaluation framework should satisfy for distributional fairness. In Section~\ref{sec:data}, we show how existing TREC test collections can be repurposed to study fairness, and Section~\ref{sec:bias} measures potential data bias to inform test collection design for fair search. A set of analyses presented in Section~\ref{sec:exp} show metric divergence between relevance and fairness, and a simple but flexible interpolation strategy for integrating relevance and fairness into a single metric for optimization and evaluation. We first provide brief background.



\section{Related Work}
IR systems face existing societal bias, such as gender \citep{chen2018investigating}, and racial \citep{noble2018algorithms} inequality. Application areas such as resume-search \cite{Zehlike2017FAIRAF} and political news-search \cite{epstein2015search, liao2013beyond, lease2018fact} demonstrate this. Two types of fairness are \textit{Individual Fairness} and \textit{Group Fairness} \cite{chen2018investigating}. Group Fairness ensures that all protected groups are treated equally, while Individual Fairness ensures that all individuals are treated equally\footnote{Fairness-measures: \url{http://www.fairness-measures.org/}}. 


Researchers have proposed different methods to tackle the bias in IR systems \citep{Zehlike2017FAIRAF, biega2018equity, ai2018unbiased, Zehlike2018ReducingDE, Yang2017MeasuringFI, Celis2018RankingWF}. These approaches include new ranking algorithms taking fairness constraints into account \citep{Celis2018RankingWF}, post-processing method for re-ranking existing systems considering both individual and group fairness \citep{Zehlike2017FAIRAF}, and evaluation of ranking systems in terms of fairness \cite{Yang2017MeasuringFI} from a group fairness perspective. IBM-360\footnote{\url{https://aif360.mybluemix.net/}} 
is an industry standard for evaluating fairness in machine learning algorithms and datasets. However, it does not include measurements for ranking systems. Recent work such as \citet{Yang2017MeasuringFI, sapiezynski2019quantifying} has begun to explore evaluating search fairness. 


\section{Evaluation Desiderata} 
\label{subsec:desiderata}



%
Motivated by \citet{lioma2017evaluation}, we propose desiderata for evaluating ranking systems that are both fair and relevant. We also include the notion of authoritativeness, especially keeping in mind modern challenges such as polarization and misinformation. 

\begin{description}
    \item[D1] \emph{Fairness}: A ranking system should return a set of documents that fairly represent different types of contents. Fairness can have different definitions. Two of them are described below.
    \item[D1.1] \emph{Equality}: A ranking system should return documents from different types in equal proportion, regardless of the distribution of content 
    \item[D1.2] \emph{Equity}: A ranking system should return documents from different types where the frequency of documents reflects the distribution of contents in real world
    \item[D2] \emph{Exposure Bias}: In the ranking output, there should not be any presentation bias for relevant documents across different types, i.e., certain types of document should not always appear before the other types \cite{biega2018equity, singh2018fairness} \ad{citations added}
    \item[D2.1] The property of Fair Exposure should hold across any $Q$ number of arbitrary queries
    \item[D3] \emph{Relevance}: A system should always return a document that is more relevant above a less relevant document 
    \item[D4] \emph{Generalizability}: At any K-th intersection of a ranking, documents should be both relevant and fairly represented
    \item[D5] \emph{Authoritativeness}: For some specific topics, a ranking system can have a deliberate authoritative bias imposed on a type of information to avoid misinformation
\end{description}

\section{Approach} 
\label{sec:approach}

In this section we describe our conceptual framework for evaluating fairness in search results. The key tenet of our approach is that documents can be organized into categories, and that IR systems should ensure some degree of balanced coverage over these categories in search results. Key concepts include:

\begin{description}
    \item [Document Categories.] we assume a single set of static document categories (e.g., topical: news vs.\ sports documents), irrespective of topic. At evaluation, we know the number of categories and labels for each document. 
    \item [Results Distribution.] The actual distribution of documents over categories in search results, notated as $R(c)$, the results distribution (as estimated) over categories.
    \item [Target Distribution.] The desired distribution of documents over categories in search results.  The target distribution may be arbitrarily specified (e.g., uniform), reflect a distributional prior, or be empirically-derived (e.g., a {\em dataset distribution}). We assume in this work that the target distribution is constant across search topics. We denote the target distribution as $Q_T(c)$, where $T$ denotes the target distribution type.
    
    \item [Estimation.] Each of the above distributions may be estimated from observed data, simply by relative frequency (maximum likelihood), or with some form of regulation or smoothing. In this work, we apply simple add-1 Laplacian smoothing\footnote{\url{https://en.wikipedia.org/wiki/Additive_smoothing}} when estimating empirical results and dataset distributions. 
    
    \item [Distributional Fairness.] We define ``fairness'' by distributional similarity: how closely the results distribution matches the target distribution. We must specify  distributional similarity. 

    \item [Integrative measures.] Beyond measuring relevance and fairness as distinct aspects of system performance, it can be useful to integrate them into a single measure. This requires specifying how metrics can be combined.

\end{description}

{\bf The Target Distribution.} What distribution should be targeted? All things being equal (i.e., lacking prior information), a uniform distribution targets balanced, equal coverage and respecting the principle of maximum entropy\footnote{\url{https://en.wikipedia.org/wiki/Principle_of_maximum_entropy}}. However, given prior information, one may specify a non-uniform target distribution.  For example, we may expect search results to respect some prior distribution. For a given dataset, we might observe a given {\em population distribution} (e.g., perhaps 65\% of documents are written in English) and want search results to be representative of this larger population. 

%


{\bf Fairness.} As noted above, we define ``fairness'' by distributional similarity between results $R(c)$ and target $Q_T(c)$ distributions. Specifically, we compute KL-divergence\footnote{\url{https://en.wikipedia.org/wiki/Kullback-Leibler_divergence}}. As discussed in the next sub-section, it is useful to have relevance and fairness on the same scale before combining them into a single measure. We thus apply min-max normalization\footnote{\url{https://en.wikipedia.org/wiki/Feature_scaling##Rescaling_(min-max_normalization)}} N[$\cdot$]. This produces a score $\in[0,1]$, but with 0 as most fair (no distributional divergence) and 1 as least fair.  For ease of interpretation and consistency with relevance metrics, we reverse the scale so F=1 is most fair.  Given a target distribution $T$, we thus compute 
    $\textnormal{Fairness} ~F_T = 1 - N[ ~KL\left( R(c) || Q_T(c)\right)~ ]$.
\textbf{Combining Relevance and Fairness.} Regardless of how relevance (R) and fairness (F) are measured, it may be useful to integrate these into a single measure. For example, F-measure interpolates between precision and recall via the harmonic mean.

In this work, we apply arithmetic and geometric means as simple interpolation methods between normalized fairness and relevance scores\cite{Ekstrand:2019:FDR:3331184.3331380, mehrotra2018towards}\ad{Added citation from the interpolation slide in the tutorial}. As with MAP vs. gMAP, the arithmetic mean is more tolerant to imbalance in inputs, whereas the geometric mean more heavily penalizes such imbalance. In general, one can specify a smoothing parameter to weight the mixture (e.g., F-measure lets one weight precision vs.\ recall, though simple, unweighted F-1 is  typically used). We leave such parameterized interpolation for future work but note the flexibility exists for balancing R and F.

As noted above, we apply min-max normalization to define fairness in range [0,1]. While relevance measures are typically also defined in that same interval, we apply consistent normalization to R as well so both R and F fully span [0,1] before they are mixed.

\textbf{Axiomatic Analysis} 
Our approach to measuring fairness is grounded in the idea of diversity. For example, "intent-aware" evaluation metrics originally developed for topical diversity \cite{agrawal2009diversifying} could be adapted to evaluating  over protected attributes for diversity. The general connection between topical diversity and fairness has also been noticed elsewhere \citet{Ekstrand:2019:FDR:3331184.3331380}. 

Our framework incorporates two different aspects of the proposed evaluation desiderata, \emph{D1} and \emph{D3}. The notion of distributional fairness can satisfy different definitions of fairness as well. As long as a target distribution can be estimated based on a particular definition of fairness, our approach can be used to score a search system. The idea of \emph{Generalizability (D4)} of metrics is also incorporated in our approach. Our method can provide insight on the fairness aspect as well as the relevance aspect of search results at any k-th intersection. Our combined metric (gmean) also enforces that a system needs to perform well both in terms of relevance and fairness. There is an implicit sense of authoritativeness (\emph{D5}) incorporated into our approach. Since one of the ways we measure fairness is to compare with the dataset distribution, a test-collection with authoritative stands for some topics encourages systems to retrieve more results from some perspectives vs.\ others. However, since we do not consider rank order here in calculating fairness, 
a system performing well on our fairness metrics can still have exposure bias (as defined in D2). 

\begin{figure*}[h!]
    \centering
    \begin{subfigure}[t]{0.49\textwidth}
        \centering
        \includegraphics[scale=0.35]{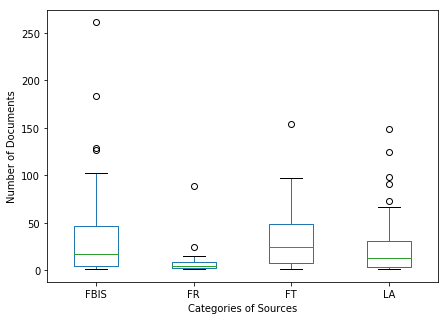}
        \caption{TREC-8 Adhoc: distribution over 4 newswire sources: Foreign Broadcast Info.\ Service, Fed.\ Register, Financial Times, and LA Times.}
        \label{subfig:news}
    \end{subfigure}
    \hfill
    \begin{subfigure}[t]{0.49\textwidth}
        \centering
        \includegraphics[scale=0.35]{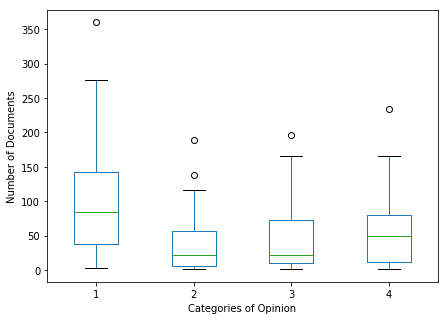}
        \caption{TREC Blogs07: distribution over 4 categories of opinionated content: 1:No Opinion, 2:Negative, 3:Mixed, and 4:Positive.}
        \label{subfig:blog}
    \end{subfigure}%
    \vspace{-1em}
    \caption{Distribution of relevant documents across topics by category for two test collections.} 
    \label{fig:EDA}
\end{figure*}

\begin{figure*}[t!]
    \begin{subfigure}[t]{0.5\textwidth}
        \centering
        \includegraphics[scale=0.43]{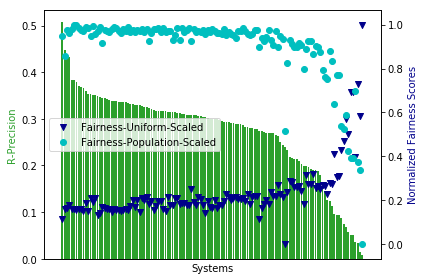}
        \caption{R-Precision vs.\ fairness scores and in TREC-8 system runs.}
        \label{subfig:compareScores-news}
    \end{subfigure}%
    \hfill
    \begin{subfigure}[t]{0.5\textwidth}
        \centering
        \includegraphics[scale=0.43]{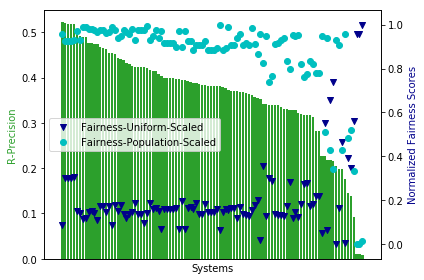}
        \caption{R-Precision vs.\ fairness scores and in Blogs07 system runs.}
        \label{subfig:compareScores-blog}
    \end{subfigure}
    \vspace{-1em}
    \caption{Correlation in system scores by metrics for relevance vs. fairness (for uniform vs.\ dataset target distributions).}
    \label{fig:Scores}
\end{figure*}

\section{Experiments}
\label{sec:exp}





Evaluation seeks to understand two over-arching questions: 1) How system performance varies under relevance vs.\ fairness (or combined) metrics; and 2) How well these metrics meet evaluation desiderata proposed in Section \ref{subsec:desiderata}. 

\label{sec:data}
\textbf{Datasets: Blogs07 and TREC-8 Adhoc}. We first describe how we adapt two existing TREC test collections to study fairness.

TREC-8 Adhoc \cite{Voorhees1999OverviewOT} considers topics 401-450 with binary relevance judgments for four Newswire sources: Financial Times, Los Angeles Times, Foreign Broadcast Information Service, and Federal Register. 129 participant rankings were obtained from TREC. This track also did not consider fairness, and we re-interpret system performance here with a new assumption that a fair ranking should provide diverse coverage across these different four news sources. 

The Blogs07 \cite{MacDonald2007OverviewOT} opinion retrieval task had participant ranking systems retrieve relevant blog posts with a given opinion for 50 topics. The collection contains binary relevance judgments and four opinion labels: no opinion, negative, mixed, and positive. 104 participant rankings were obtained from TREC\footnote{\url{http://trec.nist.gov/results/}}. While the track did not consider fairness, we re-interpret performance of those systems under a new assumption that a fair ranking should ensure diverse coverage across the four opinion categories. 

        
\subsection{Identifying Potential Test Collection Bias}
\label{sec:bias}
Our first analysis explores whether the underlying test collections are balanced or imbalanced across categories. As discussed earlier, more or less balance in relevant information across categories in an underlying test collection will likely lead to more or less balanced coverage of categories (i.e., distributional fairness) in search results. Note that since we are repurposing existing datasets to study this notion of distributional fairness, we are not disputing anything about the particular test collections under consideration, but rather describing a method by which one could design or assess a test collection over actual categories of interest for ensuring fairness.

Figures \ref{subfig:news} and \ref{subfig:blog} present the distribution of relevant documents across topics by category for each collection. As noted above, TREC-8 categories are the four newswire sources, while for Blogs07, we have four categories of opinion. 
We see that relevant information for some categories is more scarce (e.g., FR for TREC-8) or abundant (no opinion for Blogs07), so there is potential for imbalance in relevant information that could lead to imbalance in retrieved results.


\vspace{-1.5em}
\subsection{Score correlation: Relevance vs.\ Fairness}

We next aim to understand the degree to which relevance and fairness (for uniform or dataset target distributions) are correlated.  We hypothesize low correlation, which would motivate measuring both metrics and potentially optimizing retrieval results for some combination thereof. We measure R-Precision as our relevance metric due to its robustness when there are few relevant documents for a given category, per the previous analysis.

Figure \ref{fig:Scores} shows R-Precision vs.\ fairness scores for participating systems in TREC-8 and Blogs07 tracks.  System scores are sorted by decreasing R-Prec, shown as green bars, with scores measured on the left y-axis.  For each system, we also see corresponding fairness scores for two target distributions: uniform (blue) and dataset population (cyan), as measured on the right y-axis.  

The figures confirm that systems indeed perform differently for fairness and relevance metrics. We see that R-Precision scores of systems and the uniform target fairness scores are inversely correlated. However, when we compare the R-Precision scores with the dataset population target distribution, scores are more correlated (i.e., a random sample of relevant documents would tend to be representative of the population distribution in the test collection). The more imbalanced relevant documents are by category in the collection, the more population and uniform will diverge. We also infer that ranking systems are usually optimized to reflect on the distribution of documents in the test collection itself. Figures \ref{fig:EDA} and \ref{fig:Scores} suggest that to build fair ranking systems, we should also focus on developing fair test-collections as well. 

\vspace{-1em}
\subsection{Top Systems: Relevance vs.\ Fairness}

Another way to look at the relationship between relevance and fairness is to look at which systems perform best for each metric.
%
%
Tables \ref{tab:NewsWire-Results} and \ref{tab:blogs-results} report the top-3 systems for each metric, as well as for arithmetic and geometric means which integrate both measures. We see that the highest performing systems for R-Precision have very low fairness scores, and vice versa. Naturally, this inverse relationship between relevance and fairness also influences both arithmetic and geometric means. For both tracks, we see that the top systems for relevance are largely also the top systems for the arithmetic mean, but the geometric mean penalizes low fairness more heavily and so tends to select systems that are more balanced across relevance and fairness.

Recalling our high-level evaluation desiderata (Section~\ref{subsec:desiderata}), R-Precision does not tell us much beyond \emph{D3} (Relevance). Similarly, fairness does not inform us about any other evaluation criteria except for \emph{D1} (Fairness). However, the interpolation of fairness-relevance scores helps incorporate all of \emph{D1}, \emph{D3}, and \emph{D4}. 

\begin{table}[h!]
\begin{tabular}{ccccc}
\multicolumn{1}{l}{\textbf{Systems}} & \textbf{N[R-Prec]} & \textbf{Fair} & \textbf{mean} & \textbf{gmean} \\ \hline
SN1                                  & \underline{\textbf{1.0000}}            & 0.1158                   & \underline{\textbf{0.5579}}      & 0.3403                \\
SN2                                  & 0.8800                     & 0.1578                   & 0.5189               & 0.3727                \\
SN3                                  & 0.8552                     & \textbf{0.1638}          & 0.5095               & \textbf{0.3743}       \\ \hline
SN129                                & 0.0000                     & \underline{\textbf{1.0000}}          & \textbf{0.5000}      & 0.0000                \\
SN127                                & 0.0536                     & 0.7306                   & 0.3921               & 0.1979                \\
SN125                                & \textbf{0.0868}            & 0.6916                   & 0.3892               & \textbf{0.2450}       \\ \hline
SN1                                  & \multicolumn{4}{c}{see SN1 above}\\
SN2                                  & \multicolumn{4}{c}{see SN2 above}\\
SN4                                  & 0.8472                     & \textbf{0.1766}          & 0.5119               & \textbf{0.3868}       \\ \hline
SN56                                 & 0.6000                     & \textbf{0.2514}          & 0.4257               & \underline{\textbf{0.3884}}       \\
SN4                                  & \multicolumn{4}{c}{see SN4 above}\\
SN15                                 & 0.6854                     & 0.2112                   & 0.4482               & 0.3805                \\ \hline
\end{tabular}
\caption{The top-3 scoring systems for TREC8 for each of 4 metrics: relevance (R: R-Precision), fairness (F, Section~\ref{sec:approach}), and R-F arithmetic and geometric means. N[($\cdot$)] indicates min-max normalized scores, as discussed earlier. We name Systems by rank order under R-Prec (e.g., SN1 achieves best R-Prec on TREC8 Newswire, followed by SN2, etc.). The top score for each metric is underlined.}
\label{tab:NewsWire-Results}
\vspace{-3em}
\end{table}

\begin{table}[h!]
\begin{tabular}{ccccc}
\multicolumn{1}{l}{\textbf{Systems}} & \textbf{N[R-Prec]} & \textbf{Fair} & \textbf{mean} & \multicolumn{1}{l}{\textbf{gmean}} \\ \hline
SB1                                  & \underline{\textbf{1.0000}}            & 0.0861                   & 0.5431                 & 0.2935                                      \\
SB2                                  & 0.9926                     & 0.2997                   & \textbf{0.6461}        & 0.5454                                      \\
SB3                                  & 0.9906                     & \textbf{0.3006}          & 0.6456                 & \textbf{0.5457}                             \\ \hline
SB104                                & 0.0000                     & \underline{\textbf{1.0000}}          & \textbf{0.5000}        & 0.0000                                      \\
SB103                                & \textbf{0.0035}                 & 0.9560                       & 0.4797 & \textbf{0.0579}          \\
SB102                                & 0.0035                     & 0.9560                   & 0.4797                 & \textbf{0.0579}                             \\ \hline
SB4                                  & 0.9905                     & \textbf{0.3031}          & \underline{\textbf{0.6468}}        & \textbf{0.5479}                             \\
SB5                                  & 0.9891                     & 0.3040                   & 0.6465                 & 0.5483                                      \\
SB2                                  & \multicolumn{4}{c}{see SB2 above}\\
\hline
SB5                                  & 0.9891                     & \textbf{0.3039}          & 0.6465                 & \underline{\textbf{0.5483}}                             \\
SB4                                  & \multicolumn{4}{c}{see SB4 above}\\
SB3                                  & \multicolumn{4}{c}{see SB3 above}\\ \hline
\end{tabular}
\caption{Blogs07 Results, akin to TREC8 results in Table~\ref{tab:NewsWire-Results}.}
\label{tab:blogs-results}
\vspace{-3em}
\end{table}

\subsection{Rank Correlation: Relevance vs.\ Fairness}

Another question is how evaluating systems based on relevance vs.\ fairness leads to different relative orderings over participant systems. To explore this, we assume a baseline ordering of participant systems based on R-Precision, then consider how system rankings based on fairness measure differ, as measured by Kendall's $\tau$. 



Table \ref{tab:rankCorreletation} show that the rank-correlation across these metrics is quite low. The top 2 rows consider 2 target distributions: uniform and population (i.e., ground truth in dataset). The bottom 2 rows consider ranking induced by mean and gmean between uniform target and R-Precision. This adds further evidence to our earlier results in showing that evaluating systems by relevance vs.\ fairness leads to quite different results in our assessment of IR systems. 
Moreover, this highlights the need to consider 
both relevance and fairness based metrics in designing and optimizing algorithms. 



\begin{table}
\begin{tabular}{|l|c|c|c|}
\hline
\textbf{Ranking Metric} & \textbf{TREC8} & \textbf{Blogs07} \\ \hline
$F_U$: Fair$_{\textnormal{target=uniform}}$        & 0.01623                  &  -0.08028              \\ \hline
$F_P$: Fair$_{\textnormal{target=population}}$         & 0.03997                  & -0.05489               \\ \hline
$mean(F_U, \textnormal{R-Prec)}$  & 0.08503   & 0.03958   \\ \hline
$gmean(F_U, \textnormal{R-Prec})$  & 0.08503   & 0.12957   \\ \hline
\end{tabular}
\caption{Kendall's $\tau$ rank correlation over participant systems when ranked by relevance metric only (R-Precision) vs.\ ranking by fairness or relevance-fairness interpolation.}
\label{tab:rankCorreletation}
\vspace{-3em}
\end{table}

\section{Conclusion}

We defined a notion of {\em distributional fairness} and provide a conceptual framework for evaluating of search results based on it. As part of this work, we formulated a set of axioms which an ideal evaluation framework should satisfy for distributional fairness. We showed how existing TREC test collections can be repurposed to study fairness, and we measured potential data bias to inform test collection design for fair search. A set of analyses showed metric divergence between relevance and fairness, and we described a simple but flexible interpolation strategy for integrating relevance and fairness into a single metric for optimization and evaluation. 

{\bf Limitations.} We have repurposed existing TREC test collections to study fairness, but it would be better to avoid surrogate data. While we have defined fairness on a set-basis, our distributional approach can be easily extended to estimate the results distribution based on rank information, assigning greater weight to categories observed at higher ranks, addressing the exposure bias (D2) desiderata currently missed. Our min-max normalization simplifies metric combination but causes scores to change based on which systems are being compared, so this should also be revisited. \ad{Added the citation for feedback loop} There is also a scope of feedback loops reinforcing biases in search systems\cite{ensign2017runaway}. A future direction would be to expand the evaluation desiderata to have a measure for feedback loops. 
\bibliographystyle{ACM-Reference-Format}
\bibliography{References}

%


\end{document}